\documentclass[floatfix, aps, amsmath, nofootinbib, twocolumn, 10pt]{revtex4}
\usepackage{listings}
\usepackage{graphicx}

\usepackage{bm}
\usepackage{rotating}
\usepackage{array}
\usepackage{amsmath}
\usepackage{amssymb} 
\usepackage{mathrsfs} 
\usepackage{cancel}
\usepackage{subfig}
\usepackage{float}
\usepackage{caption}

\def\({\left(}
\def\){\right)}
\def\[{\left[}
\def\]{\right]}

\def\e{\begin{equation}}
\def\q{\end{equation}}
\def\m{\begin{eqnarray}}
\def\n{\end{eqnarray}}

\begin{document}
\title{Frozen Neutron Stars in Four-Dimensional Non-polynomial Gravities}
\author{Chen Tan$^{1, 2}$}
\author{Yong-Qiang Wang$^{1, 2}$}
\thanks{Corresponding author: {yqwang@lzu.edu.cn}}
\affiliation{$^{1}$Lanzhou Center for Theoretical Physics, Key Laboratory of Theoretical Physics of Gansu Province, 
	School of Physical Science and Technology, Lanzhou University, Lanzhou 730000, China}
\affiliation{$^{2}$Institute of Theoretical Physics $\&$ Research Center of Gravitation, Lanzhou University, Lanzhou 730000, China}

\date{\today}

\begin{abstract}
This paper investigates the structure and properties of neutron stars in four-dimensional non-polynomial gravities. Solving the modified Tolman–Oppenheimer–Volkoff equations for three different equations of state (BSk19, SLy4, AP4), we confirm that neutron star solutions remain in existence. As the modification parameter $\alpha$ increases, neutron stars grow in both radius and mass. We find that, when the parameter $\alpha$ is sufficiently large, a frozen state emerges at the end of the neutron-star sequence. In this state, the metric functions approach zero extremely close to the stellar surface, forming a critical horizon, making it nearly indistinguishable from a black hole to an external observer. Such a frozen neutron star constitutes a universal endpoint of the neutron-star sequence in this theory, independent of the choice of the equation of state. Based on our results and current observational constraints, we derive bounds on the modification parameter $\alpha$ and show that frozen neutron stars remain allowed in the bounds. 
\end{abstract} 

\maketitle


\section{Introduction}
\label{sec:intro}
Neutron stars, the ultracompact remnants of massive stellar collapse, offer one of the most extreme environments in the universe for probing fundamental physics. Their interiors reach supranuclear densities \cite{Oertel:2016bki}, while their strong gravitational fields \cite{LIGOScientific:2017zic,Shao:2022koz} enable stringent tests of gravity beyond the weak-field limit accessible in the Solar System. A wide range of modified gravity scenarios have been explored in the context of neutron stars, including scalar-tensor \cite{Silva:2014fca,Damour:1993hw,Pani:2014jra,Brown:2022kbw,Barranco:2021auj,Cisterna:2016vdx,Kase:2020qvz,Cisterna:2015yla,Boumaza:2023wuc}, vector-tensor \cite{Eling:2007xh,Kase:2020yhw,Ji:2024aeg}, and higher-derivative theories \cite{Charmousis:2021npl,Doneva:2023kkz,Liu:2024wvw,Saavedra:2024fzy}. Additionally, prior research \cite{Tan:2025jcg} utilized the Bardeen and Hayward nonlinear electrodynamics models as phenomenological tools, which further extended the study of neutron star structure and properties to investigations of singularity-free gravity theories. 

It is widely believed that spacetime singularities signal
the limitations of General Relativity (GR) and that a complete theory of gravity must be singularity-free. Therefore, they must be resolved within a theory of quantum gravity. Against this backdrop, a major line of research focuses on models for regular black holes (RBHs), with early work having established the Bardeen and Hayward spacetimes \cite{Bardeen:1968, Hayward:2005gi} which remain the most notable and simplest models. Subsequent research has speculated that the source capable of yielding the Bardeen black hole solution from Einstein's field equations are magnetic monopoles in nonlinear electrodynamics \cite{Ayon-Beato:1998hmi, Ayon-Beato:2000mjt}. It is popular in the literature to consider RBHs as solutions to theories that involve GR coupled to nonlinear electrodynamics. On the one hand, magnetic monopoles have not been experimentally observed; on the other hand, these models themselves suffer from numerous pathological features \cite{Bronnikov:2000vy,Bronnikov:2000yz,Fan:2016hvf,DeFelice:2024seu,Huang:2025uhv} and, in fact, still encompass all the singular solutions present in GR (these solutions are recovered when the nonlinear electromagnetic field vanishes). This has prompted researchers to continuously explore new and more natural as well as reasonable methods for constructing regular black holes.

A very recent development has seen a new approach emerge. In this approach, RBHs appear as exact solutions in theories that incorporate infinite towers of higher-curvature corrections to GR \cite{Bueno:2024dgm}. This mechanism is particularly compelling because such higher-curvature corrections are a common prediction of most quantum gravity frameworks \cite{tHooft:1974toh,Goroff:1985th,Zwiebach:1985uq,Gross:1986mw,Grisaru:1986vi,Sakharov:1967nyk,Endlich:2017tqa,Eichhorn:2020mte,Borissova:2022clg}. The method used in Ref.~\cite{Bueno:2024dgm} relies on quasi-topological gravities (QT). Boson stars, as a type of compact object, have been studied within this theory \cite{Ma:2024olw}, revealing a novel phenomenon — the existence of frozen states — which also emerges in both boson stars \cite{Wang:2023tdz,Yue:2023sep,Zhao:2025hdg,Chicaiza-Medina:2025wul} and neutron stars \cite{Tan:2025jcg} within the Bardeen and Hayward models. At frozen states within the star’s critical horizon, the metric component $-g_{tt}$ approaches zero. For a distant observer, such stellar configurations may exhibit properties analogous to those of an extremal black hole. The characteristics of these solutions are consistent with those of a frozen star, which is a theoretical model first arising from Oppenheimer and Snyder's analysis of gravitational collapse in black hole formation  \cite{Oppenheimer:1939ue}, and later formally named by Y. Zel'dovich and I. Novikov  \cite{zeldovichbookorpaper}. When observed from a distant perspective, the collapse of an ultra-compact object appears to occur over an extended period, creating that the star is frozen at their own gravitational radius \cite{Ruffini:1971bza}. This provides a solid foundation for further exploring. Meanwhile, the most recent research, utilizing the mechanism in Ref.~\cite{Bueno:2024dgm}, has been further generalized to construct a singularity-removing four-dimensional gravitational theory \cite{Bueno:2025zaj}, namely Four-Dimensional Non-polynomial Gravity. It does not involve extra dimensions and can effectively simplify related studies, providing a powerful tool for further exploring the possible novel structures and properties of neutron stars within singularity-free theories of gravity. 

In this paper, we investigate the structure and properties of neutron stars within the framework of four-dimensional non-polynomial gravity theories, with the aim of further exploring the structural and physical characteristics of neutron stars in singularity-free gravitational theories. Our results show that, under a relatively large modification parameter $\alpha$ and sufficiently high central density, these systems can exhibit frozen states in which the metric function approaches zero at the surface, making them virtually indistinguishable from black holes—analogous to those discussed in previous studies \cite{Tan:2025jcg,Wang:2023tdz,Yue:2023sep,Zhao:2025hdg,Chicaiza-Medina:2025wul,Brihaye:2025dlq}. This suggests that the frozen state could be a significant phenomenon within singularity-free gravitational theories, potentially rooted in deeper physical origins.

The structure of this paper is as follows. In Sec.~\ref{fame}, we introduce the four-dimensional non-polynomial gravities and derive the modified Tolman–Oppenheimer–Volkoff equations. In Sec.~\ref{nu}, we present the numerical results of neutron star solutions, analyze the effects of the modification parameter $\alpha$, investigate the emergence of frozen states, and examine the mass-radius relations under observational constraints. We conclude and discuss in Sec.~\ref{su}.

\section{Framework}
\label{fame}
\subsection{The Model}
We consider four-dimensional theories constructed from infinite towers of non-polynomial QTs \cite{Bueno:2025zaj}:
\begin{equation} 
\label{act1} 
S=\int d^{4}x \sqrt{-g} \left \{\frac{c^{3}}{16\pi G}\left [R+\sum^{\infty}_{n=2}\alpha_{n}Z_{(n)}\right ] \right \}.
\end{equation}
The theory is well defined on spherical backgrounds
is that their spherical sector is equivalent to two-dimensional Horndeski theories, just like the usual polynomial QTs \cite{Bueno:2024eig,Bueno:2024zsx}
\begin{equation} 
\label{act2} 
S_{2d}=\frac{c^3}{2G}\int  d^{2}x\sqrt{-\gamma}\mathcal{L}_{2d}(\gamma_{\mu\nu},\varphi) ,
\end{equation}
where
\begin{align}
\label{L2d} 
\mathcal{L}_{2d} = G_{2}(\varphi,X) - \Box\varphi \, G_{3}(\varphi,X) + G_{4}(\varphi,X) R^{2d}\\
- 2G_{4,X}(\varphi,X) \left[ (\Box\varphi)^{2} - \nabla_{\mu}\nabla_{\nu}\varphi \, \nabla^{\mu}\nabla^{\nu}\varphi \right],
\end{align}
having defined the functions $G_i(\varphi,X)$ as follows:
\begin{align}
G_2(\varphi,X) &=\varphi^2\left[3h(\psi)-2\psi h'(\psi)\right]\,,\\
G_3(\varphi,X) &=2\varphi h'(\psi)\,,\\
G_4(\varphi,X) &=-\frac{1}{2}\varphi^2\psi\int\mathrm{d}\psi\,\psi^{-2}h'(\psi)\,,
\end{align}
with
\begin{equation}
\psi=\frac{1-X}{\varphi^2}\,,\quad X=\nabla_\mu\varphi\nabla^\mu\varphi.
\end{equation}
All the information
about the 4-dimensional theory Eq.~(\ref{act1}) is now encoded
in the characteristic function $h(\psi)$
\begin{equation} 
h(\psi)\equiv\psi+\sum^{\infty}_{n=2}(2-n)\alpha_{n} \psi^n.
\end{equation}
The complete set of gravitational equations of motion for four-dimensional non-polynomial QTs
take the form:
\begin{equation} 
\mathcal{E}_{ab} = \frac{8\pi G}{c^4} T_{ab}
\end{equation}
The two-dimensional components ($t,r$) unique non-zero components of the equations of motion on are given by
\begin{equation} 
\label{eab}
\mathcal{E}_{\mu\nu} = \frac{2G_{3}}{\varphi^{2}}g_{\mu[\nu}\nabla_{\beta]}\partial^{\beta}\varphi - \frac{G_{2}}{\varphi^{2}}g_{\mu\nu}
\end{equation}
where $\mu,\nu$ denote two-dimensional components ($t,r$).

\subsection{Modified Tolman–Oppenheimer–Volkoff Equations}
Let us fix the gauge $\varphi = r$ and set
\begin{equation} 
\label{gab}
 ds_{\gamma}^{2} = -N^{2}(r) f(r)c^2dt^{2} + \frac{dr^{2}}{f(r)}, \quad \varphi = r.
\end{equation}
We treat the matter as a perfect fluid with energy-momentum tensor 
\begin{equation}
\label{tab}
T_{\mu\nu}=(\rho c^2+p)U_{\mu}U_{\nu}+pg_{\mu\nu},
\end{equation}
where $\rho c^2$ and $p$ are the energy density and pressure of the matter. And normalized to $U^{\mu}U_{\mu}=-1$, it becomes
\begin{equation}
U_{\mu}=(N(r)\sqrt{f(r)}, 0). 
\end{equation} 
Substituting Eq.~(\ref{gab}) and Eq.~(\ref{tab}) into Eq.~(\ref{eab}) yields the following  differential equations from $\mathcal{E}_{tt}$,$\mathcal{E}_{rr}$
\begin{align}
\label{ett}
\mathcal{E}_{tt}=&\partial_{r}(r^3h(\psi))=\frac{8\pi G r^2\rho(r)}{c^2},\\ 
\label{err}
\mathcal{E}_{rr}=&-\frac{3h(\psi)}{f(r)} + \frac{2\psi\, h'(\psi)}{f(r)}\notag\\
&+ \frac{h'(\psi)}{r} \left( \frac{f'(r)}{f(r)} + \frac{2N'(r)}{N(r)} \right)=\frac{8\pi G\, p(r)}{f(r)c^4} 
\end{align}

From considering Eq.~(\ref{ett}), it is natural to simplify the calculations by defining the mass function
\begin{equation}
m(r)=\frac{c^2}{2G}r^3h(\psi)=4\pi \int^{r}_{0}\rho(x)x^2dx.
\end{equation}
The energy-momentum conservation $\bigtriangledown _{\mu}T^{\mu\nu}=0$, and it gives
\begin{equation}
\frac{1}{2} \big(p(r) + \rho(r)c^2\big) 
\left( \frac{f'(r)}{f(r)} + \frac{2\, N'(r)}{N(r)} \right) 
+ p'(r)=0.
\end{equation}
By combining this result with Eq.~(\ref{err}), we eliminate $\left( \frac{f'(r)}{f(r)} + \frac{2\, N'(r)}{N(r)} \right) $ to derive the modified Tolman-Oppenheimer-Volkoff (TOV) equations
\begin{align}
\label{tov}
&m'(r)=4\pi r^2 \rho(r),\\
&p'(r)=\notag\\&-\frac{r \left(c^{2} \rho(r) + p(r)\right) \left(3 h(\psi) - 2 \psi h'(\psi) +\frac{8 \pi G p(r)}{c^4}\right)}{2 f(r) h'(\psi)}.
\end{align}
Once the parameters $(2-n)\alpha_n$ are specified, the explicit forms of the characteristic function $h(\psi)$ and the metric function $f(r)$ are determined by solving the $\mathcal{E}_{tt}$ equation together with the relation $\psi=\frac{1-X}{\varphi^2}=\frac{1 - f(r)}{r^2}$. When $(2-n)\alpha_n=0$, these equations reduce to the original form derived by Tolman, Oppenheimer and Volkoff  \cite{Tolman:1939jz, Oppenheimer:1939ne}.

\subsection{Choices of $(2-n)\alpha_n$}
\begin{figure*}[]
\begin{center}
\subfloat{\includegraphics[width=0.755\textwidth]{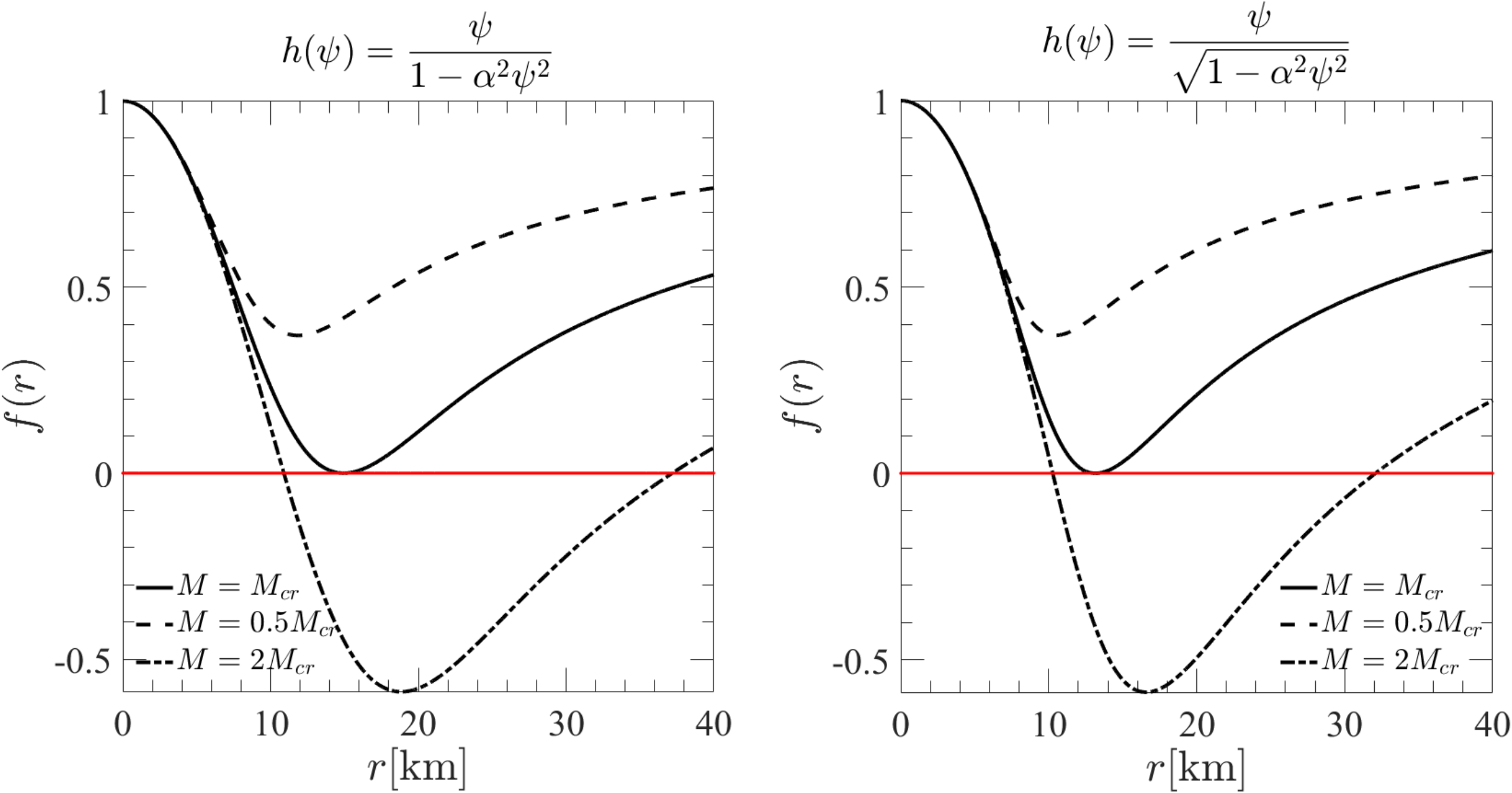}}
\end{center}
\captionsetup{justification=raggedright}
\caption{The black curves $f(r)$ correspond to different configurations of the mass function $m$: the dashed curve ($m = 0.5M_{\text{cr}}$) represents a horizonless regular spacetime; the solid curve ($m = M_{\text{cr}}$), an extremal black hole with a single horizon; and the dot-dashed curve ($m = 2M_{\text{cr}}$), a regular black hole with two horizons.}
\label{EBH}
\end{figure*}

Different non-polynomial QTs theories, distinguished by their respective choices of the parameters $\alpha_n$, each possess distinct black hole solutions. When $(2-n)\alpha_n = 0$, the modified TOV equations reduce to their general relativity counterpart. We focus on specific models for concreteness to derive and solve the modified TOV equations, those that yield simple expressions and enable analytic calculations. In odd dimensions, a suitable choices within these models give rise to $D$-dimensional versions of the Hayward black hole \cite{Hayward:2005gi}. However, in even dimensions, the existence of such solutions is prohibited because the term of order $n = D/2$ is topological. $\mathcal{Z}_{(2)}$ is the Gauss–Bonnet invariant. In $D = 4$, it contributes nothing to the equations of motion, which results in the absence of the quadratic term in $h(\psi)$. Alternative choices of couplings which yields simple uniparametric family of solutions in $D = 4$ corresponds to the characteristic polynomial \cite{Bueno:2024dgm}
\begin{align}
&h(\psi)=\frac{\psi}{1-\alpha^{2}\psi^{2}},\ \alpha > 0,\\
&h(\psi)=\frac{\psi}{\sqrt{1-\alpha^{2}\psi^{2}}},\ \alpha > 0,
\end{align}
which entails choosing
\begin{align}
\label{dalpha}
&(2-n)\alpha_{n}=\frac{(1-(-1)^n)}{2}\alpha^{n-1},\\
\label{dalpha2}
 &(2-n)\alpha_{n}=\frac{(1-(-1)^n)\Gamma\left(\frac{n}{2}\right)}{2\sqrt{\pi}\Gamma\left(\frac{n+1}{2}\right)}\alpha^{n-1}.
\end{align}
For this, the metric function $1/g_{rr}=f(r)$ reads
\begin{align}
f(r) &= 1 - \frac{4G r^2 m(r)}{c^2 r^3 + \sqrt{c^4 r^6 + 16\alpha^2 G^2 m^2(r)}}\notag\\
&\overset{Gm(r)\ll rc^2}{\sim} 1-\frac{2Gm(r)}{c^2r}(1-\frac{4\alpha^2G^2m^2(r)}{c^4r^6}+...),
\\
f(r)&= 1 - \frac{2 G r^2 m(r)}{\sqrt{c^4 r^6 + 4 \alpha^2 G^2 m^2(r)}}\notag \\
&\overset{Gm(r) \ll rc^2}{\sim} 1-\frac{2Gm(r)}{c^2r}(1-\frac{2\alpha^2G^2m^2(r)}{c^4r^6}+...).
\end{align}
When $m(r)$ is a constant $m$, these solutions reduce to vacuum solutions. The vacuum solutions describes extremal black holes (EBHs) with a single horizon (as shown in Fig.~\ref{EBH}) when $m = M_{cr}$.
\begin{align}
M_{cr}= \frac{5^{5/4} }{8}\frac{c^2}{G} \sqrt{\alpha},\ 
M_{cr}= \frac{3^{3/4}}{2^{3/2}} \frac{c^2}{G}\sqrt{\alpha}.
\end{align}

From their weak-field expansion, we can readily see that for a finite $m(r)$, at $r\to \infty$ in vacuum, $m(r)$ is precisely $M_{\text{ADM}}$ of the system, as determined from the coefficient of the $1/r$ term in the expansion. Therefore, the mass of a neutron star with a well-defined boundary can be determined
\begin{equation} 
M\equiv m(R)=4\pi \int^{R}_{0}\rho(x)x^2dx. 
\end{equation}

For Mercury's orbit, the deviation from flat spacetime predicted by General Relativity (GR) is approximately $5 \times 10^{-8}$. Even imposing a stringent upper limit of $10^{-20}$ on modified-gravity effects to ensure no observable deviation in Mercury's perihelion precession results in a comparatively weak constraint on the parameter $\alpha < 10^{22}~[\rm{m}^2]$. In fact,this imposes a more stringent constraint than measurements of the Earth's surface gravitational acceleration.

\section{Numerical Calculation}
\label{nu}

 All throughout this paper, numerical solutions to the initial value problem are obtained with an adaptive 4th-order Runge-Kutta method. 
The modified TOV equations will be solved from the center
at $r=r_{\delta}$ to the surface ($p=\rho=0$) of the star at $r=R$, satisfying
the boundary conditions:
\begin{equation}
m(r_{\delta}) = \frac{4}{3}\pi r_{\delta}^3\rho _c,\  \rho c^2(r_{\delta}) = \rho _c c^2, p(r_{\delta})=p_c
\end{equation}
where $\rho_{c}$ and $p_{c}$ are the central density and pressure respectively, and $r_{\delta}$ corresponds to the core radius which we take to be $r_{\delta}=10^{-4}$\rm{m} $\ll R$. We have checked that all of our results are independent of the choice of $r_{\delta}$ provided this is a very small number relative to the NS radius. We numerically solved the modified TOV equations using three different equations of states (EOS) to investigate neutron star physical properties under four-dimensional non-polynomial gravities. The stiffness of the EoS models employed (BSk19 \cite{Potekhin:2013qqa}, SLy4 \cite{Douchin:2001sv,Haensel:2004nu}, AP4 \cite{Akmal:1997ft, Gungor:2011vq}) increases progressively. We proceed values of central density between $\rho_c = 2.5\times 10^{14}[\rm{g/cm^{3}} ]$ and the causal-limit $\rho_{c}=\rho_{max}$ (where $v_s=c$ as Tab.~\ref{rhomax} show). However, not all densities in this interval are allowed in all cases. For sufficiently large $\alpha$, there exists a critical central density $\rho_{\rm{cr}}$ beyond which no physically meaningful numerical solution can be obtained. As $\rho_{\rm{c}}$ approaches $\rho_{\rm{cr}}$, the minimum of $1/g_{rr}$ tends to zero, and the corresponding solution asymptotically approaches a black hole — the frozen state mentioned earlier. A detailed discussion of this frozen state will be presented in Sec.~\ref{fs}.

\begin{table}[htbp] 
\renewcommand\arraystretch{1.5}
\captionsetup{justification=raggedright}
\caption{Maximum allowed density from causality-constrained equations of state.}
\label{rhomax}
\begin{tabular}{ |c | c  c  c  c  | } 
\hline 
EOS & BSk19 & SLy4 & AP4 &  \\ 
\hline 
$\rho_{max}$& $3.3814\times 10^{15}$ & $3.0075\times 10^{15}$ & $1.7069\times10^{15}$\ &$[\rm{g/cm^3}]$ \\
\hline 
\end{tabular}
\end{table}

\subsection{The Effect of $\alpha$} 
\begin{figure*}
\begin{center}
\subfloat{\includegraphics[width=0.75\textwidth]{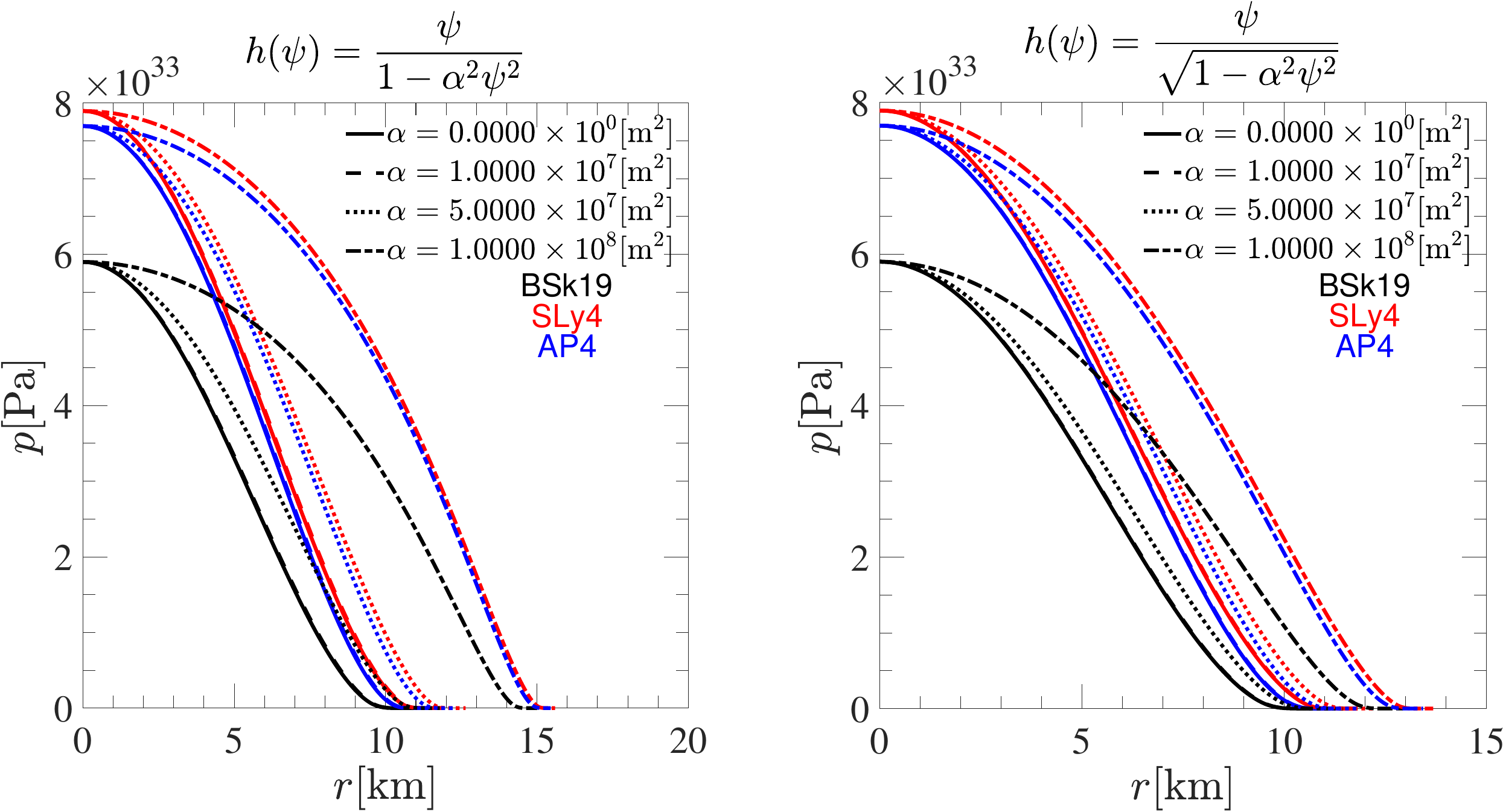}}
\end{center}
\captionsetup{justification=raggedright}
\caption{
The left and right subplots show the radial pressure profiles for different characteristic functions ($\frac{\psi}{1-\alpha^2\psi^2}$ and $\frac{\psi}{\sqrt{1-\alpha^2\psi^2}}$) at a fixed central density $0.8\times 10^{15} \rm[g/cm^3]$, plotted for various values of the modification parameter $\alpha$. Results from the BSk19(black), SLy4(red), and AP4(blue) equations of state are compared.}
\label{pr}
\end{figure*}

The radial pressure distribution was obtained directly
by solving the modified Tolman-Oppenheimer-Volkoff
(TOV) equations.
Fig.~\ref{pr} presents the radial pressure profiles of neutron star matters in $\rho_c = 0.8 \times 10^{15}\rm{[g/cm^3]}$

For both adopted characteristic function forms $h(\psi)$, an increase in the modification parameter $\alpha$ results in a larger neutron star radius, accompanied by a shallower radial pressure gradient (Fig.~\ref{pr}). A more pronounced modification is observed for $\frac{\psi}{1-\alpha^{2}\psi^{2}}$ compared to $\frac{\psi}{\sqrt{1-\alpha^{2}\psi^{2}}}$ at identical $\alpha$, as implied by Eq.~(\ref{dalpha}) and Eq.~(\ref{dalpha2}).

It can be observed that, despite the shared ability to construct regular black holes, the resulting pressure profile of neutron stars in four-dimensional non-polynomial gravities differs markedly from that in Bardeen/Hayward models. Notably, it does not develop the characteristic ``filled hard candy like” structure seen in those models. A similar discrepancy has been noted in earlier research on boson stars \cite{Ma:2024olw}. This implies that the structure of neutron stars in singularity-free gravity can be significantly influenced by the theory's physical origin and the specific modification schemes employed.

\begin{figure*}[]
\begin{center}
\subfloat{\includegraphics[width=0.75\textwidth]{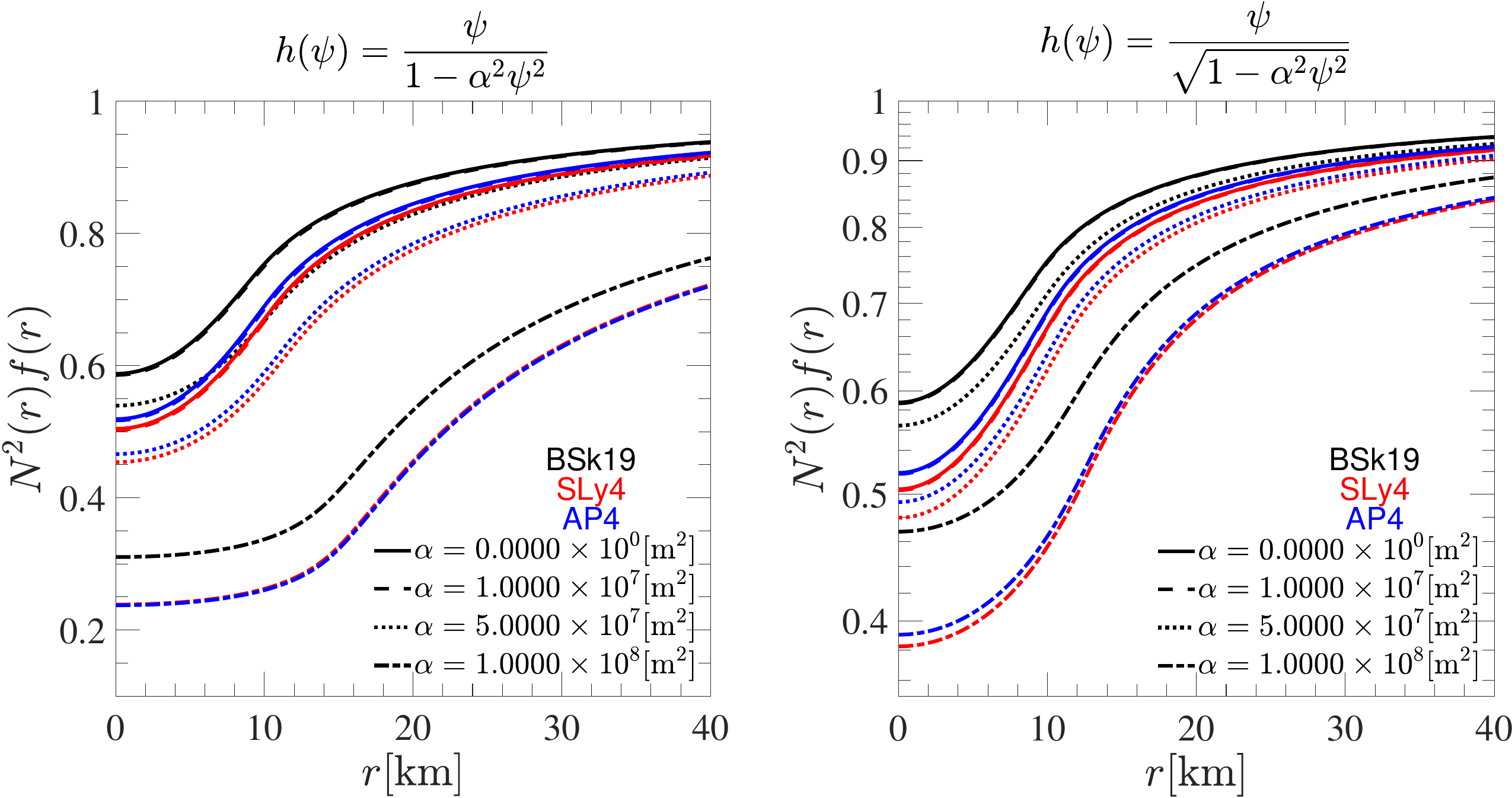}}
\end{center}
\captionsetup{justification=raggedright}
\caption{The left and right subplots show $N^2(r)f(r)$ ($-g_{tt}$) for different characteristic functions ($\frac{\psi}{1-\alpha^2\psi^2}$ and $\frac{\psi}{\sqrt{1-\alpha^2\psi^2}}$) at a fixed central density $0.8\times 10^{15} \rm[g/cm^3]$, plotted for various values of the modification parameter $\alpha$. Results from the BSk19(black), SLy4(red), and AP4(blue) equations of state are compared. }
\label{gtt}
\end{figure*}

\begin{figure*}[]
\begin{center}
\subfloat{\includegraphics[width=0.75\textwidth]{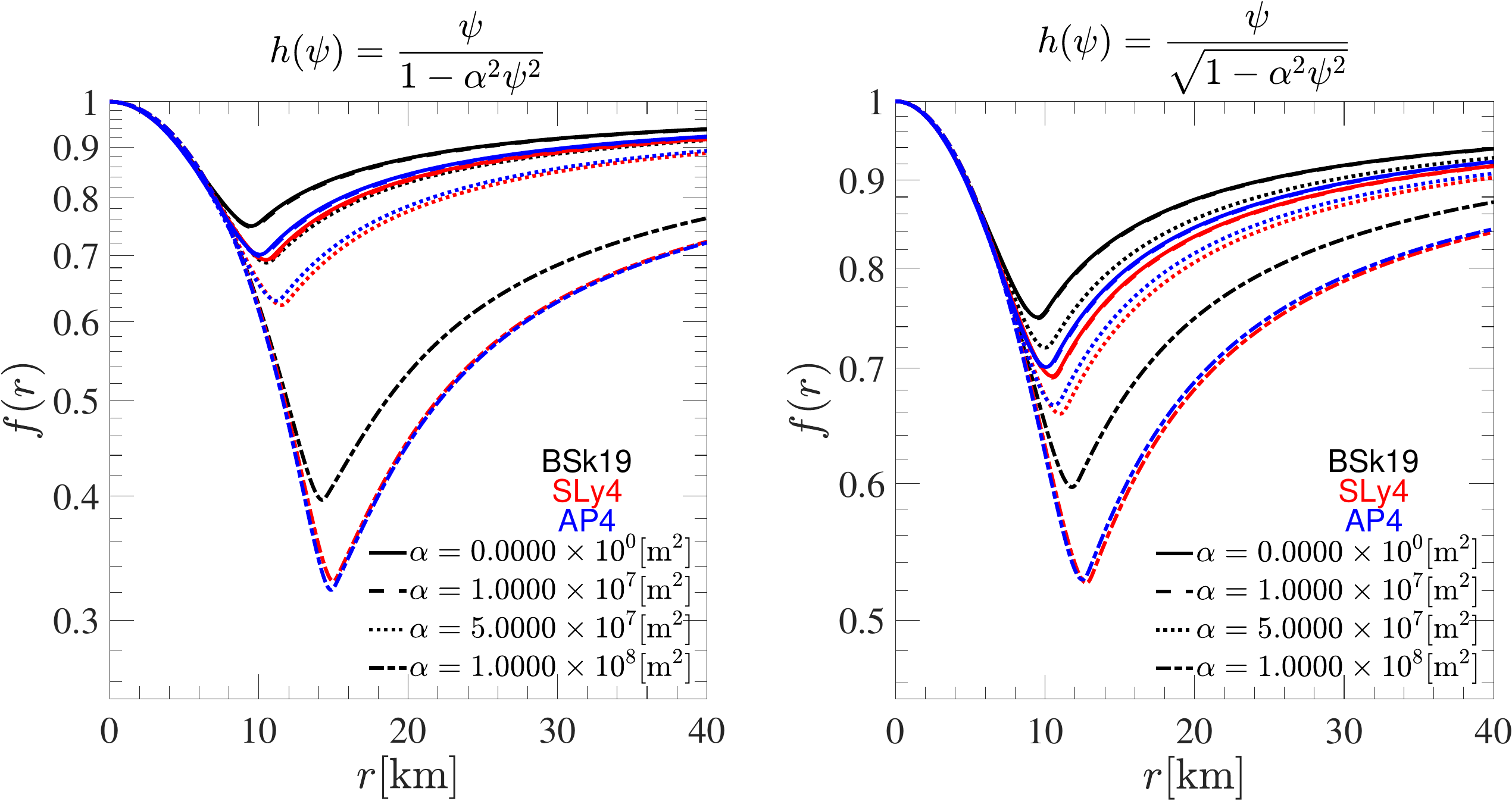}}
\end{center}
\captionsetup{justification=raggedright}
\caption{The left and right subplots show $f(r)$ ($1/g_{rr}$) for different characteristic functions ($\frac{\psi}{1-\alpha^2\psi^2}$ and $\frac{\psi}{\sqrt{1-\alpha^2\psi^2}}$) at a fixed central density $0.8\times 10^{15} \rm[g/cm^3]$, plotted for various values of the modification parameter $\alpha$. Results from the BSk19(black), SLy4(red), and AP4(blue) equations of state are compared. }
\label{grr}
\end{figure*}

In Fig.~\ref{gtt} and Fig.~\ref{grr} ,we further observe that the minima of the metric functions $N^2(r)f(r)$ ($-g_{tt}$) and $f(r)$ ($1/g_{rr}$) decrease, with the minimum of  $f(r)$ shifting outward. This behavior is consistent with the growth in stellar radius and total mass, as illustrated in Fig.~\ref{pr}.   

\begin{figure*}[]
\begin{center}
\subfloat{\includegraphics[width=0.75\textwidth]{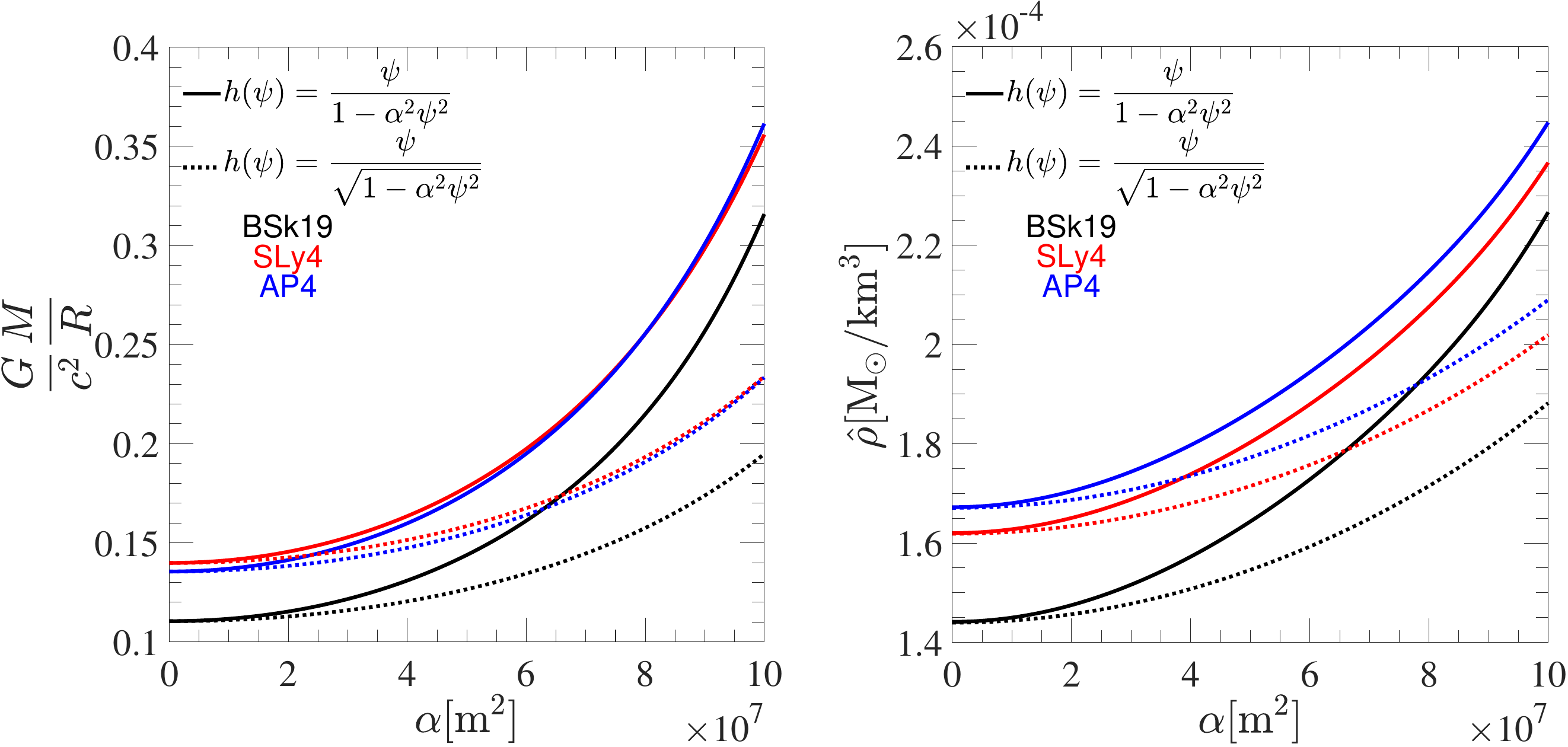}}
\end{center}
\captionsetup{justification=raggedright}
\caption{The dependence of the compactness $\mathcal{C}$ and average density $\bar{\rho}$ on the modification parameter $\alpha$ is depicted in the left and right subplots for different characteristic functions ($\frac{\psi}{1-\alpha^2\psi^2}$ and $\frac{\psi}{\sqrt{1-\alpha^2\psi^2}}$). Results from the BSk19(black), SLy4(red), and AP4(blue) equations of state are compared.}
\label{corho}
\end{figure*}

Under two distinct characteristic functions $h(\psi)$, both the compactness $\mathcal{C}$ and the average density $\bar{\rho}$ are found to increase with the parameter $\alpha$ (as Fig.~\ref{corho}). This trend, as derived from the following expression:
\begin{align}
\label{c}
\mathcal{C}=\frac{G M}{c^2 R}= \frac{G}{c^2}\frac{M}{4/3\pi R^3}\frac{4/3\pi R^3}{R}\propto\hat{\rho}R^2. 
\end{align}

 Due to the positive correlation between pressure and density in the equation of state, this behavior is already foreshadowed by the radial pressure profiles presented in Fig.~\ref{pr}, where the variations with $\alpha$ indicate a concurrent increase in both $\hat{\rho}$ and the stellar radius. Consequently, the overall compactness of the system increases with $\alpha$.

\subsection{Frozen States}\label{fs}

We find that for large $\alpha$, when the central density $\rho_{c}$ exceeds a certain critical  $\rho_{cr}$, numerical solutions cease to yield physically acceptable results. 

\begin{figure}[ht]
\begin{center}
\subfloat{\includegraphics[width=0.75\columnwidth]{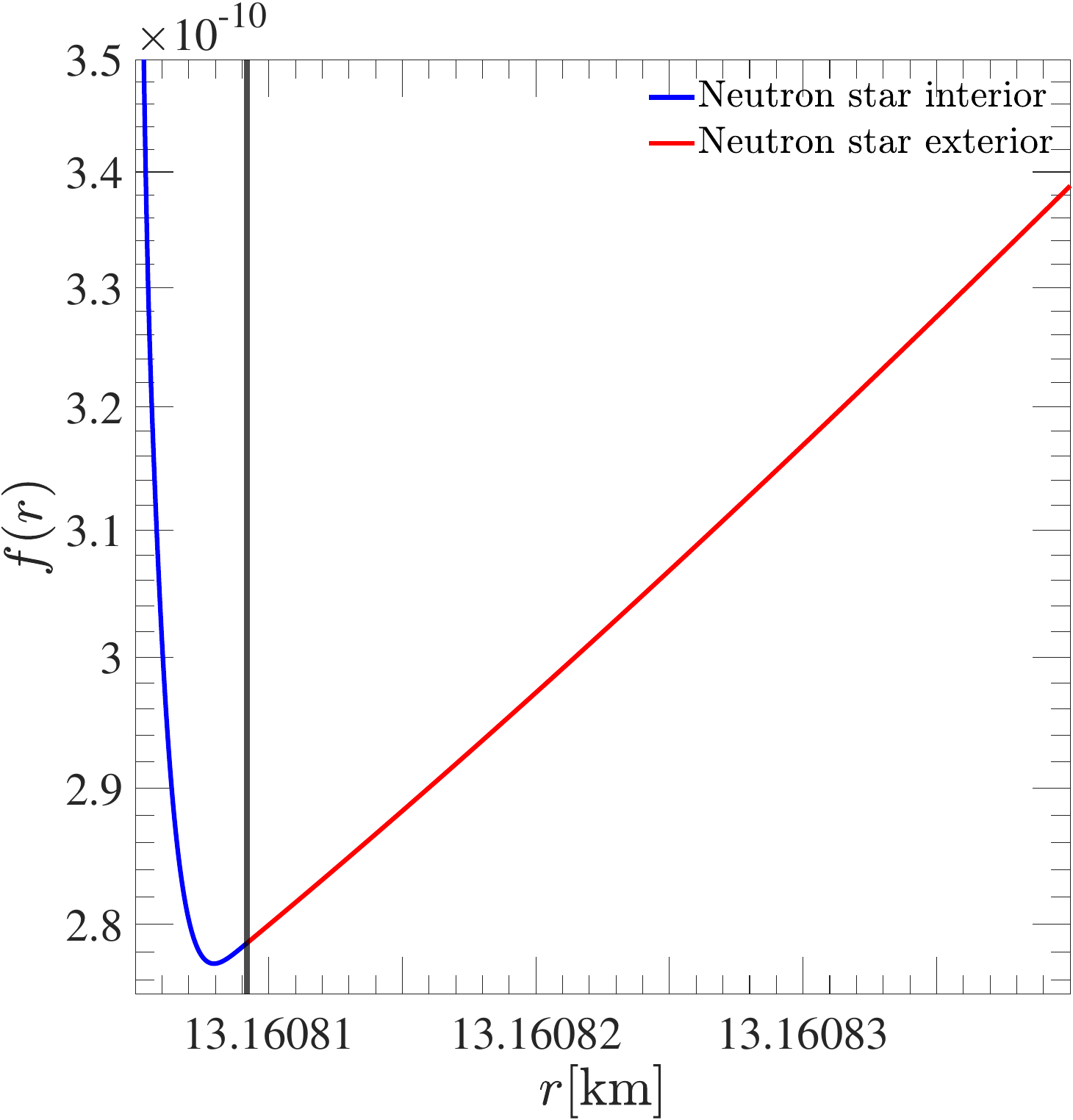}}
\end{center}
\captionsetup{justification=raggedright}
\caption{The figure illustrates $f(r)$ of BSk19 at $\rho_{c}$ under characteristic function $\frac{\psi}{\sqrt{1-\alpha^2\psi^2}}$, with the neutron star interior shown in blue and the exterior in red. A black vertical line marks the boundary of the neutron star. It is clear that the critical horizon, although very close to the surface of the neutron star, does not actually coincide with it. }
\label{crsu}
\end{figure}

\begin{figure*}[]
\begin{center}
\subfloat{\includegraphics[width=0.75\textwidth]{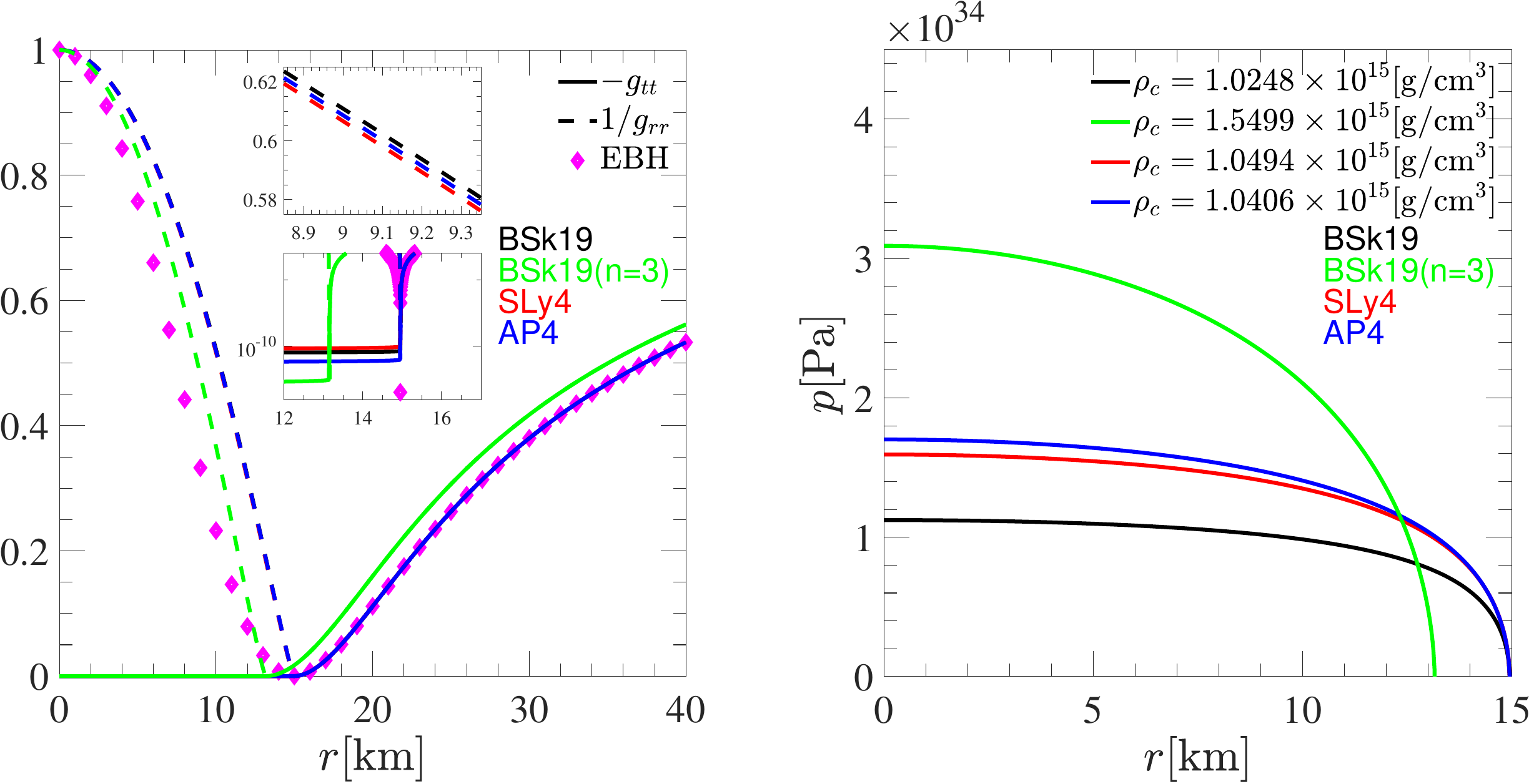}}
\end{center}
\captionsetup{justification=raggedright}
\caption{Under the characteristic function $\frac{\psi}{1-a^2\psi^2}$, the left subplots shows the metric function in the frozen state, and the right subplot shows the radial pressure distribution in the frozen state.Results from the BSk19(black), SLy4(red), and AP4(blue) equations of state are compared. And the green curve takes BSk19 as an example, illustrating the frozen state under the case of a finite truncation (n=3).}
\label{fz1}
\end{figure*}

\begin{figure*}[]
\begin{center}
\subfloat{\includegraphics[width=0.75\textwidth]{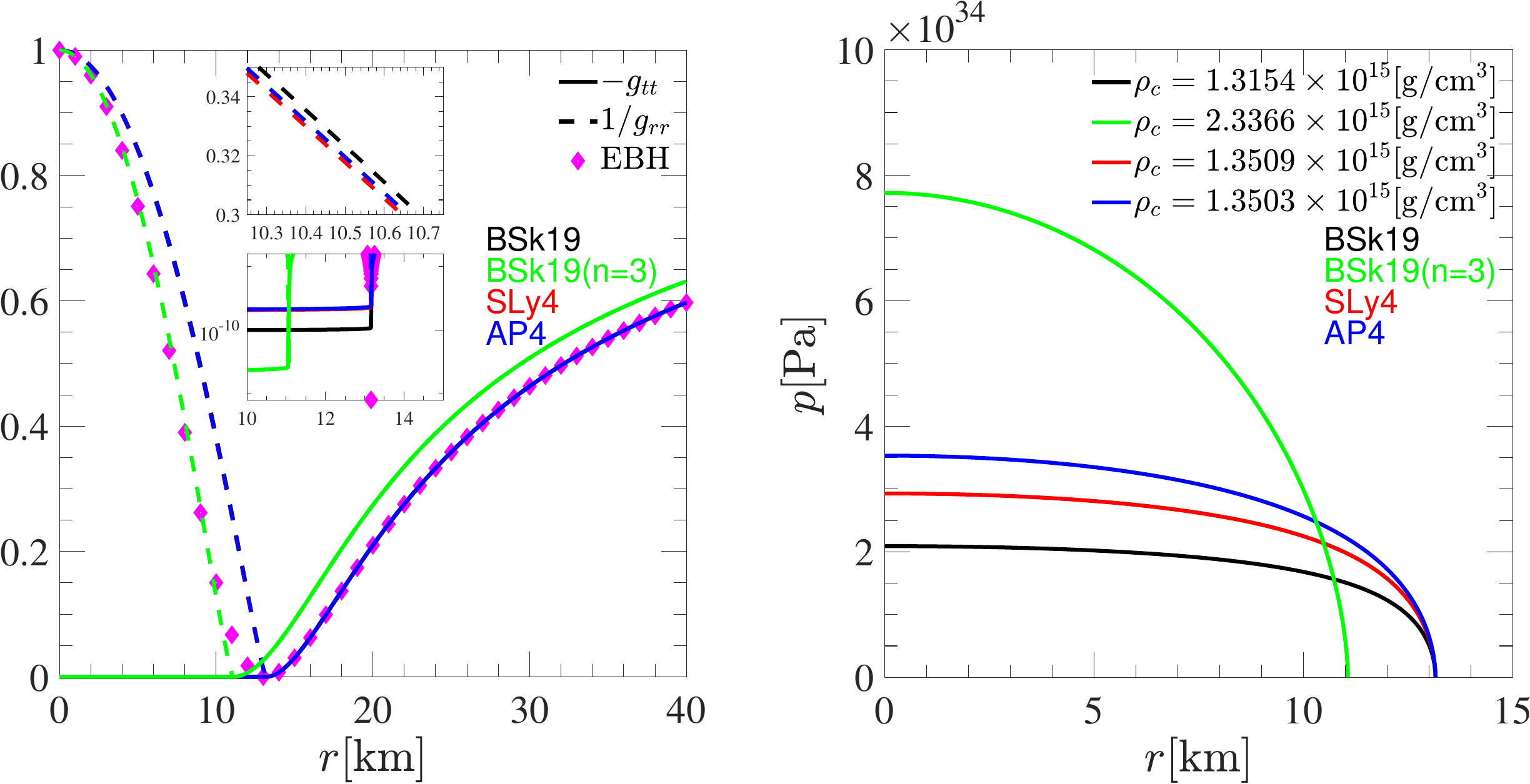}}
\end{center}
\captionsetup{justification=raggedright}
\caption{Under the characteristic function $\frac{\psi}{\sqrt{1-a^2\psi^2}}$, the left subplots shows the metric function in the frozen state, and the right subplot shows the radial pressure distribution in the frozen state. Results from the BSk19(black), SLy4(red), and AP4(blue) equations of state are compared. And the green curve takes BSk19 as an example, illustrating the frozen state under the case of a finite truncation (n=3). }
\label{fz2}
\end{figure*}

When reaching this critical density $\rho_{cr}$, under the selection of two different characteristic functions $h(\psi)$, the neutron star enters the same frozen state as described in the literature \cite{Tan:2025jcg,Wang:2023tdz,Yue:2023sep,Zhao:2025hdg,Chicaiza-Medina:2025wul,Brihaye:2025dlq}. A remarkable feature of this state is that the minimum value of $1/g_{rr}=f(r)$ is extremely close to zero, and the product $-g_{tt}=N^{2}(r)f(r)$ inside the location of this minimum is also extremely close to zero, as Fig.~\ref{fz1} and Fig.~\ref{fz2} shown. This minimum location has been referred to as the critical horizon in previous studies. It is worth noting, however, that at this stage, the critical horizon only approaches the surface of the neutron star very closely and does not exactly coincide with it. That is, the critical horizon does not encompass all the matter inside the neutron star; beyond it, there remains an extremely thin layer of matter (see Fig.~\ref{crsu}).
From an external perspective, its metric functions are nearly indistinguishable from those of an extreme black hole, with virtually identical mass and radius. Furthermore, it can be observed that the properties of the frozen state are almost independent of the choice of the equation of state. All the equations of state presented in this study lead to the frozen state, with only very minor differences in the metric functions within this state. The nature of the equation of state primarily influences the critical density value at which the frozen state is reached (Tab.~\ref{eos1e} and Tab.~\ref{eos2e}), as well as the permissible range of $\alpha$ (Tab.~\ref{eosalpha}) that allows entry into the frozen state within the density range considered in this work. For the two different characteristic functions $\frac{\psi}{1-\alpha^{2}\psi^{2}}$ and $\frac{\psi}{\sqrt{1-\alpha^{2}\psi^{2}}}$, the former induces stronger modifications at the same $\alpha$, resulting in a lower critical density requirement for the same $\alpha$ and the same equation of state.

Further analysis reveals that the formation of the frozen state does not require summing the characteristic function to infinite order, i.e., the complete removal of the singularity. In fact, the frozen state already emerges at finite orders---as demonstrated here for $n = 3$
\begin{equation} 
h(\psi)=\psi+\alpha_{3}\psi^3.
\end{equation}
For this characteristic function, the metric takes the form
\begin{equation}
f(r)= 1 + \frac{c^2 r^4}{3^{1/3} \cdot X^{1/3}} - \frac{3^{1/3}X^{1/3}}{3\alpha_3 c^2},
\end{equation}
with
\begin{equation}
X = 9\alpha_3^2 c^4 G r^3 m(r) + \sqrt{3} \sqrt{\alpha_3^3 c^8 r^6 \bigl(c^4 r^6 + 27 \alpha_3 G^2 m^2(r) \bigr)}.
\end{equation}
where
\begin{align}
\alpha_{3}&=\alpha^2\ in  \  \frac{(1-(-1)^n)}{2}\alpha^{n-1},\\
\alpha_{3}&=\frac{1}{2}\alpha^2\  in \ \frac{(1-(-1)^n)\Gamma\left(\frac{n}{2}\right)}{2\sqrt{\pi}\Gamma\left(\frac{n+1}{2}\right)}\alpha^{n-1} .
\end{align}

In the $n = 3$ case, the modification strength---which quantifies the ability to regulate the singularity---is weaker than in the $n \to \infty$ \cite{Bueno:2025zaj}. Consequently, for a given coupling $\alpha$ and equation of state, a higher central density is required to reach the frozen state,as shown by the example of BSk19 in Fig.~\ref{fz1} and Fig.~\ref{fz2}. This behavior is consistent with earlier findings in boson star models \cite{Ma:2024olw}. Moreover, the fact that the frozen state can be realized even under finite-order truncation offers insight into the occurrence of analogous phenomena in 4D Einstein--Gauss--Bonnet gravity \cite{Charmousis:2021npl,Saavedra:2024fzy}.

\begin{table}[htbp] 
\renewcommand\arraystretch{1.5}
\captionsetup{justification=raggedright}
\caption{Critical densities $\rho_{cr}$ of different equations of state under the characteristic function $\frac{\psi}{1-\alpha^2\psi^2}$, across varying correction parameters $\alpha$.}
\label{eos1e}
\begin{tabular}{ |c |  c  c c  | } 
\hline 
$\alpha$  & $5. 0000\times 10^{7}$ & $1. 0000\times 10^{8}$ &$[\rm{m^2}]$ \\ 
\hline 
BSk19&  $2.2769\times 10^{15}$ & $1.0248\times 10^{15}$ &$[\rm{g/cm^3}]$ \\
SLy4&  $2.3657\times 10^{15}$ & $1.0494\times 10^{15}$ & $[\rm{g/cm^3}]$ \\
AP4 &  $--$ & $1.0406\times 10^{15}$ &$[\rm{g/cm^3}]$ \\
\hline 
\end{tabular}
\end{table}

\begin{table}[htbp] 
\renewcommand\arraystretch{1.5}
\captionsetup{justification=raggedright}
\caption{Critical densities $\rho_{cr}$ of different equations of state under the characteristic function $\frac{\psi}{\sqrt{1-\alpha^2\psi^2}}$, across varying correction parameters $\alpha$.}
\label{eos2e}
\begin{tabular}{ |c | c  c c  | } 
\hline 
$\alpha$ & $5.0000\times 10^{7}$ & $1.0000\times 10^{8}$ &$[\rm{m^2}]$ \\ 
\hline 
BSk19 & $2.9298\times 10^{15}$ & $1.3154\times 10^{15}$ &$[\rm{g/cm^3}]$ \\
SLy4 & $--$ & $1.3509\times 10^{15}$ & $[\rm{g/cm^3}]$ \\
AP4 & $--$ & $1.3503\times 10^{15}$ & $[\rm{g/cm^3}]$ \\
\hline 
\end{tabular}
\end{table}

\begin{table}[htbp] 
\renewcommand\arraystretch{1.5}
\captionsetup{justification=raggedright}
\caption{The minimum $\alpha_{\min}$ required for different equation of states under  causal constraints to enter the frozen state.}

\label{eosalpha}
\begin{tabular}{ |c | c  c  c  c  | } 
\hline 
EOS & BSk19 & SLy4 & AP4 &  \\ 
\hline 
$\alpha_{min}(\frac{\psi}{1-\alpha^{2}\psi^{2}})$& $3.3769\times 10^{7}$ & $3.9328\times 10^{7}$ & $6.1494\times10^{7}$\ &$[\rm{m^2}]$ \\
\hline 
$\alpha_{min}(\frac{\psi}{\sqrt{1-\alpha^{2}\psi^{2}}})$& $4.4319\times 10^{7}$ & $5.1419\times 10^{7}$ & $8.2817\times10^{7}$\ &$[\rm{m^2}]$ \\
\hline 
\end{tabular}
\end{table}

\subsection{Mass-Radius Relation}
\begin{figure*}[]
\begin{center}
\subfloat{\includegraphics[width=0.75\textwidth]{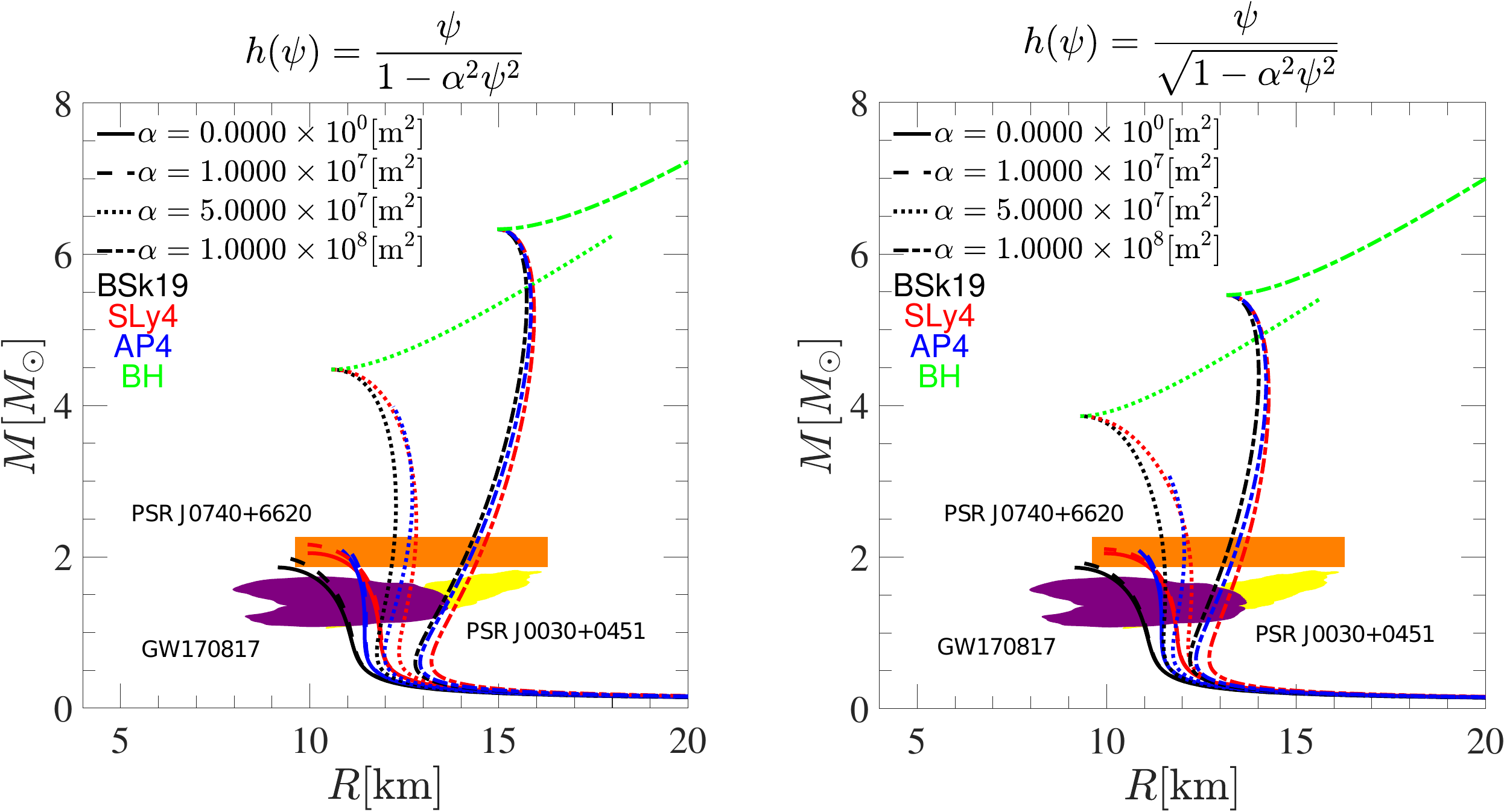}}
\end{center}
\captionsetup{justification=raggedright}
\caption{The left and right subplots show $M$-$R$ relation for different characteristic functions ($\frac{\psi}{1-\alpha^2\psi^2}$ and $\frac{\psi}{\sqrt{1-\alpha^2\psi^2}}$) , plotted for various values of the modification parameter $\alpha$. Results from the BSk19(black), SLy4(red), and AP4(blue) equations of state are compared. The black hole sequence in four-dimensional non-polynomial gravities is marked in green. Different color bands show the $2\sigma$ constraints from GW170817(purple), PSR J0740+6620(orange), PSR J0030+0451(yellow)}
\label{mr}
\end{figure*}

The mass-radius ($M$-$R$) relation is a crucial issue in neutron star research and serves as the most direct relation for constraining theories. In Fig.~\ref{mr}, we present the modifications to the $M$-$R$ relation induced by different modification parameters $\alpha$ under two distinct characteristic functions.It can be seen that as $\alpha$ increases, the $M$-$R$ relation undergoes significant deformation: the radius of the neutron star solutions increases, and the maximum mass also rises. Within the range of the parameter $\alpha$ that permits the formation of frozen neutron stars, such objects emerge at the leftmost endpoint of the mass-radius curve. We find no physically meaningful numerical solutions beyond the critical density, which results in the truncation of the mass-radius curve at the point of frozen star formation. Notably, at this truncation point, the mass gap between the neutron star sequence and the black hole sequence is nearly closed. This finding is consistent with the modified Buchdahl limits predicted by this theory \cite{Bueno:2025tli}.

Meanwhile, the color bands in the figure, representing observational constraints on the neutron star mass-radius curves from pulsar PSR J0030+0451 \cite{Riley:2019yda,Miller:2019cac}, PSR J0740+6620 \cite{Miller:2021qha}, and the gravitational wave event GW170817 \cite{LIGOScientific:2018cki}, impose limits on the modification parameter $\alpha$. For characteristic function $\frac{\psi}{1-\alpha^2\psi^2}$ and $\frac{\psi}{\sqrt{1-\alpha^2\psi^2}}$ , within its constrained range of $\alpha$ (Tab.~\ref{eosc1} and Tab.~\ref{eosc2}), the generation of frozen stars is permitted for all three equations of state currently considered. Within the parameter space of the model permitting the formation of frozen neutron stars, the maximum mass attainable by the neutron star sequence  exceeds that of the currently known observational samples. If the causality constraint on the maximum density of the equation of state were relaxed, frozen neutron stars with lower masses could be obtained at smaller $\alpha$. We impose a conservative causal-limit cutoff on the equation of state to ensure the robustness of our findings. Furthermore, it can be observed that the softer equation of state BSk19—previously ruled out by constraints from PSR J0740+6620 \cite{Miller:2021qha}—reenters the observationally allowed region when gravitational corrections are taken into account.

\begin{figure*}[]
\begin{center}
\subfloat{\includegraphics[width=0.75\textwidth]{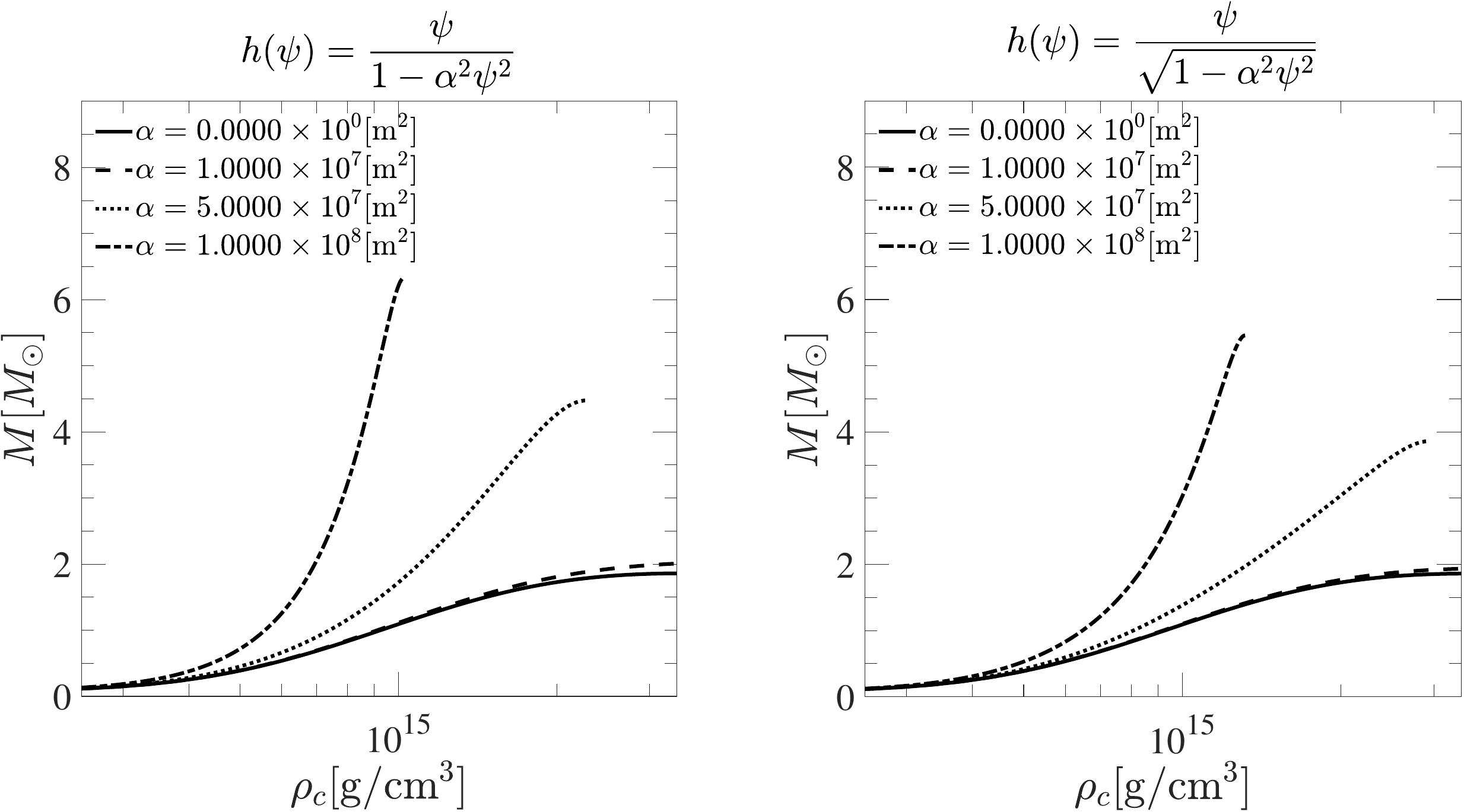}}
\end{center}
\captionsetup{justification=raggedright}
\caption{The left and right subplots show $M$-$\rho_{c}$ relation of BSk19 for different characteristic functions ($\frac{\psi}{1-\alpha^2\psi^2}$ and $\frac{\psi}{\sqrt{1-\alpha^2\psi^2}}$), plotted for various values of the modification parameter $\alpha$. The behavior of SLy4 and AP4 is similar to that of BSk19, which is taken as an example here.}
\label{rho_m}
\end{figure*}

Under causal-limit cutoff, frozen stars lie on the branch where $\partial M/\partial \rho_{c}>0$ (for example Fig.~\ref{rho_m}), and are therefore likely to be stable. While a detailed stability analysis requires further study, investigations into similar scenarios in 4D Einstein-Gauss-Bonnet gravity provide confidence in their stability \cite{Saavedra:2024fzy}.

\begin{table}[htbp] 
\renewcommand\arraystretch{1.5}
\captionsetup{justification=raggedright}
\caption{Under characteristic function $\frac{\psi}{1-\alpha^2\psi^2}$, for different equations of state, the upper and lower bounds of $\alpha$ constrained by PSR J0030+0451 \cite{Riley:2019yda,Miller:2019cac}, PSR J0740+6620 \cite{Miller:2021qha}, and GW170817 \cite{LIGOScientific:2018cki}.}
\label{eosc1}
\begin{tabular}{ |c | c  c  c  c  | } 
\hline 
EOS & BSk19 & SLy4 & AP4 &  \\ 
\hline 
$\alpha_{lower}$& $0.9\times 10^{7}$ & $0.0\times 10^{0}$ & $0.0\times10^{0}$\ &$[\rm{m^2}]$ \\
\hline 
$\alpha_{upper}$& $0.9\times 10^{8}$ & $0.8\times 10^{8}$ & $0.9\times10^{8}$\ &$[\rm{m^2}]$ \\
\hline 
\end{tabular}
\end{table}

\begin{table}[htbp] 
\renewcommand\arraystretch{1.5}
\captionsetup{justification=raggedright}
\caption{Under characteristic function $\frac{\psi}{\sqrt{1-\alpha^2\psi^2}}$, for different equations of state, the upper and lower bounds of $\alpha$ constrained by PSR J0030+0451 \cite{Riley:2019yda,Miller:2019cac}, PSR J0740+6620 \cite{Miller:2021qha}, and GW170817 \cite{LIGOScientific:2018cki}.}
\label{eosc2}
\begin{tabular}{ |c | c  c  c  c  | } 
\hline 
EOS & BSk19 & SLy4 & AP4 &  \\ 
\hline 
$\alpha_{lower}$& $1.0\times 10^{7}$ & $0.0\times 10^{0}$ & $0.0\times10^{0}$\ &$[\rm{m^2}]$ \\
\hline 
$\alpha_{upper}$& $1.3\times 10^{8}$ & $1.1\times 10^{8}$ & $1.2\times10^{8}$\ &$[\rm{m^2}]$ \\
\hline 
\end{tabular}
\end{table}

\section{Summary and Discussion}
\label{su}

In this work, we have systematically investigated the structure and properties of neutron stars within the framework of four-dimensional non-polynomial gravities \cite{Bueno:2025zaj}---a class of singularity-free theories constructed from infinite towers of higher-curvature corrections. By solving the modified Tolman--Oppenheimer--Volkoff equations for three representative equations of state, we have demonstrated that neutron star solutions persist in this theory.

Our study reveals that as the modification parameter $\alpha$ increases, neutron stars grow in both radius and mass, accompanied by shallower radial pressure gradients and increases in both the compactness and the average density. The functional form of the characteristic function $h(\psi)$ influences the strength of these modifications, with $\frac{\psi}{(1 - \alpha^2 \psi^2)}$ inducing stronger effects than $\frac{\psi}{\sqrt{1 - \alpha^2 \psi^2}}$ for the same $\alpha$.

Most significantly, for sufficiently large $\alpha$ and central densities beyond a critical value $\rho_{cr}$, neutron stars enter a ``frozen state''. In this regime, the metric functions $1/g_{rr}$ and $g_{tt}$ approach zero extremely close to the stellar surface, forming a critical horizon. To a distant observer, such objects are nearly indistinguishable from extremal black holes. The frozen state appears universally across all three EOS models, with only minor differences in critical densities and allowed $\alpha$ ranges, suggesting that frozen stars represent a generic endpoint of the neutron star sequence in this theory. We have further found that frozen states can already emerge at finite orders (for example $n = 3$), without requiring the full infinite tower of higher-curvature correction. The mass--radius curves shift toward larger radii and masses with increasing $\alpha$, with frozen neutron stars residing at the leftmost tip of these curves, where the mass gap between neutron stars and black holes nearly closes. Considering the constraints on the neutron star mass-radius curves from pulsar PSR J0030+0451, PSR J0740+6620, and the gravitational wave event GW170817, the existence of such frozen neutron stars has not been ruled out. This leaves an interesting window for future research.

While our study establishes the existence and basic properties of frozen neutron stars in four-dimensional non-polynomial gravities, several key questions remain open. A full stability analysis is required to determine whether these objects represent stable equilibrium configurations. While frozen stars are near-perfect observational mimickers of black holes, they might be distinguished through high-precision probes of their tidal deformability, quasi-normal mode spectrum, and accretion flow dynamics. Substantial work remains to fully develop these discriminators and their observational implications. The emergence of frozen states in both boson stars and neutron stars \cite{Tan:2025jcg,Wang:2023tdz,Yue:2023sep,Zhao:2025hdg,Chicaiza-Medina:2025wul,Brihaye:2025dlq} across distinct models may potentially rooted in deeper physical origins. Further investigation is required to determine whether this represents a generic feature of singularity‑free gravitational theories.

\begin{acknowledgments}
This work is supported by the National Key Research and Development Program of China (Grant No. 2022YFC2204101 and 2020YFC2201503) and the National Natural Science Foundation of China (Grant No. 12275110 and No. 12247101). 
\end{acknowledgments}

\end{document}